\documentclass[12pt]{article}
\usepackage{graphicx}
\begin{document}
\def\ie{{\it i.e.}}
\def\eg{{\it e.g.}}
\def\qv{{\it q.v.}}
\def\ib{{\it ibid.}}
\def\bc{\begin{center}}
\def\ec{\end{center}}
\def\bq{\begin{quotation}}
\def\eq{\end{quotation}}
\def\bit{\begin{itemize}}
\def\eit{\end{itemize}}
\def\benum{\begin{enumerate}}
\def\eenum{\end{enumerate}}
\def\be{\begin{equation}}
\def\ee{\end{equation}}
\def\bearr{\begin{eqnarray}}
\def\eearr{\end{eqnarray}}
\def\bdm{\begin{displaymath}}
\def\edm{\end{displaymath}}
\def\bfl{\begin{flushleft}}
\def\efl{\end{flushleft}}
\def\bfr{\begin{flushright}}
\def\efr{\end{flushright}}
\def\del{\partial}
\def\goes{\longrightarrow}
\def\epem{e^+e^-}
\def\MR{M_\Phi}
\def\VR{\Lambda_\Phi}
\bc
{\Large\bf  
On Distinguishing Radions from Higgs Bosons} \\
\vskip 10pt
{\sl Prasanta Kumar Das $^a$, Santosh Kumar Rai} $^b$ and 
{\sl Sreerup Raychaudhuri $^b$} \\
\bigskip
$^a$~{\rm Department of Physics, Chung Yuan Christian University, \\
22 Pu-Jen, Pu-chung Li, Chung-Li (32023), Taiwan, R.O.C.}\\
{\rm Electronic address:} {\sf pdas@phys.cycu.edu.tw} \\
\bigskip
$^b$~{\rm Department of Physics, Indian Institute of Technology, 
Kanpur 208016, India.} \\
{\rm Electronic addresses:} {\sf sreerup@iitk.ac.in, skrai@iitk.ac.in} \\
\vskip 20pt
{\large\bf ABSTRACT}
\ec
\bq \noindent {\footnotesize
Radion couplings are almost identical to Higgs boson couplings, making it very
difficult to distinguish the two states when the radion mass and vacuum
expectation value are similar to those of the Higgs boson. The only real
difference lies in the fact that the coupling of radions to off-shell fermions is
proportional to the momentum rather than the mass of the fermion. This extra
contribution gets cancelled in all tree-level processes and shows up only in
loop-induced processes like $\Phi \to \gamma\gamma$ and $\Phi \to gg$. We perform
a careful calculation of the branching ratios and establish that they can prove
crucial in clearly distinguishing a radion from a Higgs boson. This claim is made
concrete by evaluating the exclusive cross-sections in a radiative process
involving elementary scalars.}
\eq
\vskip 10pt

\noindent
In recent years the two-brane model of Randall and Sundrum\cite{RS1} has
attracted a great deal of attention because it provides an elegant
solution to the thirty-odd year-old hierarchy problem of high energy
physics. The most attractive feature of the 1+4 dimensional
Randall-Sundrum (RS) model is that it explains the large hierarchy
between the electroweak scale (0.1 -- 1 TeV) and the Planck scale
($10^{16}$ TeV) in terms of an exponential damping of the gravitational
field across a small compact fifth dimension, without recourse to
unnaturally large numbers\footnote{However, it is only fair to mention
that this is achieved at the expense of a delicate fine-tuning of the
five-dimensional cosmological constant with the energy densities on two
branes at opposite ends of the fifth dimension.}. Since the hierarchy of
scales is generated by an exponential damping across the fifth dimension,
the size of this dimension requires to be just at the right value to
ensure that the hierarchy is indeed a factor of $\frac{M_{Pl}}{M_{ew}}
\sim 10^{16}$. Since the fifth dimension has the topology of a
once-folded circle, with two $D_3$-branes at the fixed points, this
amounts to fixing the distance $R_c$ between the two branes very
precisely. This, in turn, would require a mechanism to protect the radius
against large quantum corrections. The absence of such a mechanism is a
major flaw in the original braneworld model of Arkani-Hamed, Dimopoulos
and Dvali\cite{ADD}. The question was left unresolved even in the
original work\cite{RS1} of RS, but an elegant model to explain this was
given shortly afterwards by Goldberger and Wise\cite{GW2}. They used the
simple device of generating a force between the two branes which would
ensure equilibrium when the distance between them is precisely the radius
$R_c$ required to generate the required hierarchy. Because of the folded
structure of the fifth dimension, it is only necessary to generate an
attractive force between the branes --- since each brane is,
topologically speaking, on {\sl both} sides of the other, the two pulls
will balance at the equilibrium point. The attractive force is modelled
by postulating the time-honoured device of a scalar field which lives in
all five dimensions (bulk)  and has quartic self-interactions, in the
bulk, as well as in projection on the branes. It is necessary only to
tune the vacuum expectation values (vev's) of the scalar field on the two
branes to get an attractive force as required. In fact, it can be easily
shown that the potential has an extremely steep minimum at the argument
$R_c$, indicating that the hierarchy is fixed very accurately for small
oscillations of the bulk size about this minimum.

\bigskip\noindent

An important consequence of the (original)  RS model is that there exists
on the TeV-brane (which represents the observable Universe), a scalar
field $\Phi$, which is very much like a dilaton field and has been dubbed
the {\sl radion}. The RS metric, with radial fluctuations, is written in
the form
\be
ds^2 = e^{-2KT(x)\varphi} ~g_{\mu\nu}(x) dx^\mu dx^\nu - [T(x)]^2 ~d\varphi^2
\ee
where $T(x)$ is a modulus field representing dilatation of the bulk, $K$
is the bulk curvature and $\varphi$ is an angular coordinate describing
the fifth dimension. We can now show\cite{GW1} that the five-dimensional
Einstein-Hilbert action reduces, in four dimensions, to a theory with
Kaluza-Klein gravity and a scalar term
\be
S_\Phi = \int d^4x \sqrt{-|g|} ~~\frac{1}{2} \del_\mu \Phi(x) \del^\mu 
\Phi(x)
\ee
where the (massless, free) scalar field
\be
\Phi(x) = \sqrt{24M_5^3/K}~e^{-\pi K T(x)}
\ee
is the radion field. The Goldberger-Wise stabilisation mechanism, with a
bulk scalar field $B(x,\varphi)$ then creates an effective scalar
$\Phi^4$-potential for the radion field $\Phi(x)$ with a minimum at
$\langle T(x)\rangle = R_c$, where $KR_c \simeq 11.7$, the value required
to generate the electroweak hierarchy. This potential includes a radion
mass term $m_\Phi^2 \Phi^2$ where $m_\Phi$ is determined by the mass
$m_B$ and couplings of the bulk scalar field. Though $m_B$ is unknown,
arguing\cite{GW2} that $m_B$ should be of the order of the Planck scale
--- which is the only fundamental scale in the RS model --- leads to the
result that $m_\Phi$ should be of the order of the electroweak
symmetry-breaking scale, i.e. $m_\Phi \sim 100$~GeV.

\bigskip\noindent
Phenomenology of the radion field\cite{DHR1} starts with the coupling of
the radion to ordinary matter, consisting of the Standard Model (SM)
fields. This interaction which arises from the usual gravitational
coupling to matter, is the same as the coupling of a dilaton field, viz.,
\be
{\cal L}(x) = -\frac{1}{\VR} ~\Phi(x) ~\eta^{\mu\nu}T_{\mu\nu}(x)
\label{radcoup}
\ee
where $\Lambda_\Phi$ is the radion vev, corresponding to the minimum of
the radion potential on the TeV brane and $T_{\mu\nu}$ is the
energy-momentum tensor composed of SM fields. The radion Feynman rules
can, therefore, be read off from (for example), the expressions given in
Ref.\cite{HLZ} by simply substituting the radion for the dilaton and
$\VR$ for $\bar M_P$. It turns out that the couplings are rather similar
to those of Higgs fields to other SM particles, though, of course, the
radion couplings originate from the 
couplings of the bulk scalar\footnote{It may be noted that the vev, 
$\VR$ is a free parameter arising from the couplings of the bulk scalar 
and has no
relation to the parameter $\Lambda_\pi$ which is strongly constrained by
bounds on the graviton mass \cite{lambdapi}.} 
while Higgs boson couplings arise from the Standard Model sector. An
important --- and for this work, crucial --- difference arises in the
fact that there are {\it momentum-dependent} terms in the radion coupling
to matter, which are not present in the Higgs boson coupling. To see
this, we write out in full the energy-momentum tensor for a scalar field
$S(x)$, a fermion field $\psi(x)$ and a vector gauge field $V_\mu(x)$.
\bearr
\eta^{\mu\nu}~T_{\mu\nu} & = & 
-2 \left[(D^\mu S)^\dagger (D_\mu S) - 2 M_S^2 ~S^\dagger S \right]
\nonumber \\ &   &
-3i \bar{\psi} \not{\!\!D} \psi + 4 m_\psi~\bar{\psi}\psi + \frac{3i}{2}
\del^\mu \left[ \bar{\psi}\gamma_\mu\psi\right]
\nonumber \\ &   &
- M_V^2 ~V_\mu(x) V^\mu(x)
\label{tensor}
\eearr
where $D^\mu = \del^\mu + igT_a V^\mu_a$ and $T_a$ is the gauge group
generator in the appropriate representation. The explicit form of this
for the Standard Model fields is given in Ref.\cite{KCH}. This
interaction tells us that the vertex for $\Phi(p) \to \bar{\psi}(k_1) +
\psi(k_2)$ is given (in momentum space) by
\be
{\cal L}_{\Phi \psi\bar \psi} = \frac{3}{2\VR} \bar u(k_1) \left(
\not{\!k_1} + \not{\!k_2} - \frac{8}{3}m_\psi \right) u(k_2) ~\Phi(p) \ .
\label{offshell}
\ee 
If the fermions are on-shell, we can use the Dirac equation to write the above
equation as
\be
{\cal L}_{\Phi \psi\bar \psi} =
-\frac{m_\psi}{\VR} \bar u(k_1) u(k_2) \Phi(p) \ ,
\label{onshell}
\ee
which is a Yukawa coupling very reminiscent of the Higgs boson.
Obviously, if the fermions are {\it off-shell}, the coupling is
different.  An immediate consequence of the above is that radions, unlike
Higgs bosons can have significant couplings to light fermions, such as
electrons and $u,d$ quarks, if any of the fermions is off-shell and has
large energy-momentum values.

\bigskip\noindent
Once produced, a massive radion will clearly decay. Being a constituent
field of the metric tensor the radion couples, as described in Eqns.
(\ref{radcoup}) and (\ref{tensor}), to {\sl all} pairs of SM particles.
Naturally, only those decays which are kinematically allowed will occur.
For fermionic decay modes $\Phi \to f\bar{f}$ (on-shell), the partial
decay widths will be suppressed by the factor $m_f^2/\Lambda_\Phi^2$,
since the fermionic states will be on-shell and the radionic coupling
will be Higgs boson-like. Thus, we need to consider only the following
decay channels:
\bearr
\Phi & \goes & \gamma\gamma,~gg \nonumber \\ 
     & \goes & \tau^+\tau^- ,~c\bar{c},~b\bar{b} \nonumber \\ 
     & \goes & W^+W^-,~ Z^0Z^0,~H^0H^0 \nonumber \\ 
     & \goes & t\bar{t}
\nonumber
\eearr
Detailed formulae for these are given in Ref.\cite{KCH}. However, in this
last-mentioned work, the formulae for the decay widths $\Phi \goes \gamma
\gamma,~g g$ have been calculated assuming the coupling in
Eqn.(\ref{onshell}) rather than the one in Eqn.(\ref{offshell}), and are,
therefore, somewhat inaccurate. This is because loop-fermions will
obviously be off-shell and there will be consequent modifications to the
decay amplitude itself. It is worth noting that though the lighter
fermions will now couple to the scalar field through their momenta rather
than their masses, there is a helicity flip involved in the
scalar-vector-vector one-loop diagrams, which results in an amplitude
proportional to the fermion masses. As a result, it is only the top quark
loop which makes any significant contribution --- which is also the case
with Higgs bosons. The differences arise, then, solely from the extra
off-shell terms in the $\Phi t\bar t$ coupling.

\bigskip\noindent
In this work, we have calculated the one-loop-mediated decay widths
afresh, using the off-shell coupling of Eqn.(\ref{offshell}). The
relevant Feynman diagrams are listed in Figure~1. Of these, the triangle
diagrams marked (A) and (B) are similar to those responsible for the
process $H^0 \to \gamma\gamma$. The ones marked (C) and (D) arise from
non-renormalisable couplings of the radion (dilaton) to a photon and a
fermion pair, which arise at the lowest order in an effective theory of
gravity coupling to matter. It is worth mentioning at this point that the
presence of these vertices is responsible, in tree-level processes like,
for example, $e^+ e^- \to Z \Phi$ or $e^+ e^- \to \ell^+\ell^-\Phi$, for
precisely cancelling out the momentum-dependent part of the
$\ell^+\ell^-\Phi$ coupling and rendering the cross-section totally
Higgs-like. However, the situation is different inside a loop diagram, as
we shall presently argue.

\begin{center}
\includegraphics[height=3in]{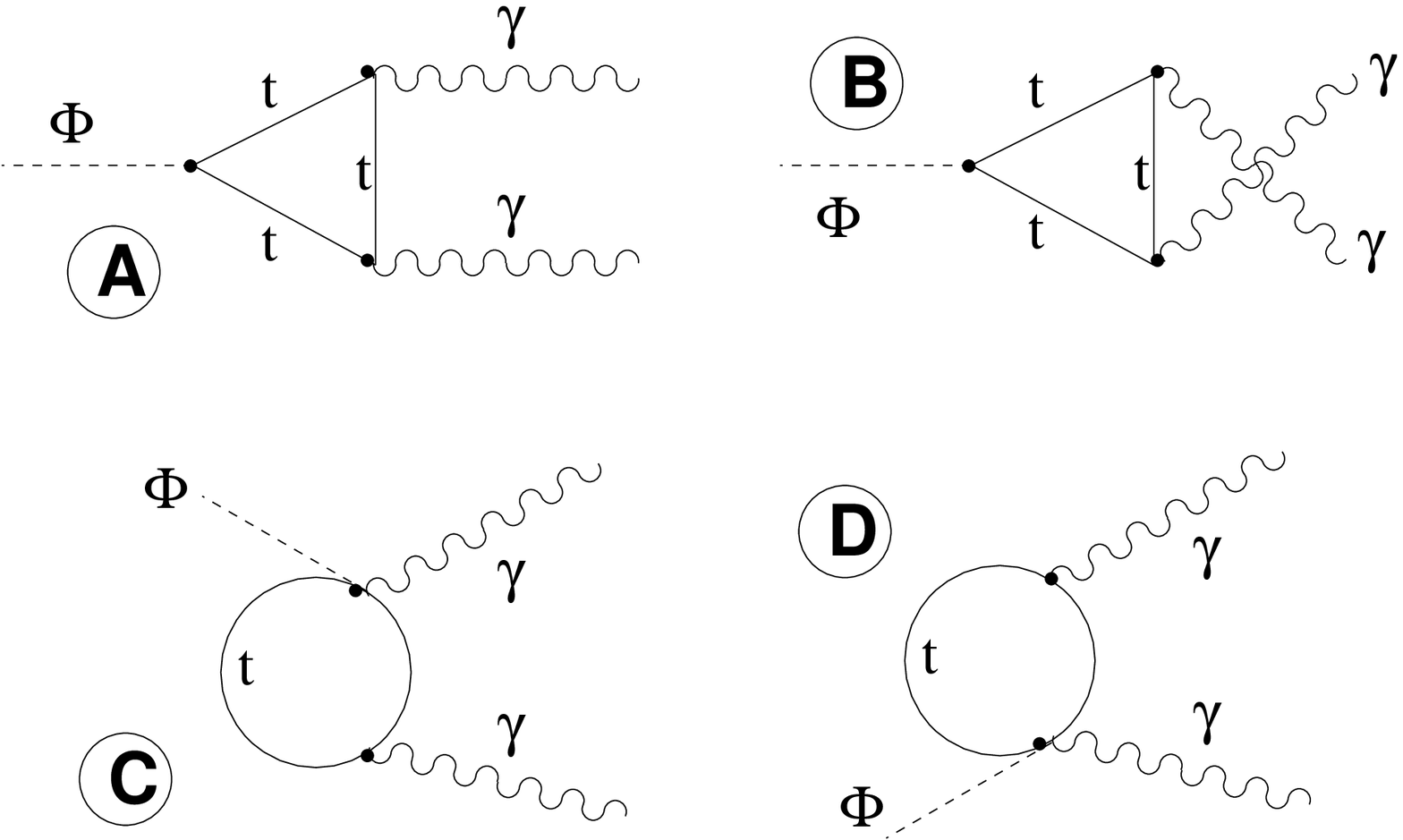}
\end{center}
\noindent {\bf Figure 1}.
{\footnotesize\it Feynman diagrams with a top quark loop contributing to the process
$\Phi \goes \gamma\gamma$ at the one-loop level.}

\bigskip\noindent 
Following the usual procedure\cite{ABJ} for calculating the amplitude for a
process like $\Phi(p) \goes \gamma(k_1)\gamma(k_2)$, we write the amplitude as
\be
{\cal M}(\Phi\to\gamma\gamma) = 
\left[ A(p^2)k_1^\nu k_2^\mu + B(p^2) \eta^{\mu\nu} \right] 
\varepsilon^*_\mu(k_1)\varepsilon^*_\nu(k_2) 
\ee
which is consistent with Lorentz-symmetry and the transverse nature of
the photon. Imposition of gauge symmetry at once leads to the Ward
identity
\be
B(p^2) = - A(p^2) k_1.k_2
\ee
which relates the (naively divergent) form factor $B(p^2)$ to the finite form 
factor $A(p^2)$, and hence acts as a regulator for the process. The amplitude
then becomes
\be
{\cal M}(\Phi\to\gamma\gamma) = 
A(p^2) \left( k_1^\nu k_2^\mu - k_1.k_2 ~\eta^{\mu\nu} \right) 
\varepsilon^*_\mu(k_1)\varepsilon^*_\nu(k_2) 
\ee
which means that it is only necessary to calculate the finite form factor
$A(p^2) = A(M_\Phi^2)$ in order to get the decay width. Since $A(p^2)$
can be calculated by evaluating the coefficients of $k_1^\nu k_2^\mu$
alone, it can now be seen, by writing down the Feynman amplitudes for the
diagrams marked (A)--(D) in Figure~1, that the contributions to $A(p^2)$
arise from those marked (A) and (B), but not from those marked (C) and
(D). Thus, the exact cancellation of momentum-dependent terms, which
renders the effective radion coupling Higgs-like in the tree-level, does
{\sl not} go through at the one-loop level. This also ensures that the
diagrams (A) and (B) in Figure~1 have residual momentum-dependent effects
and justifies corrections to the partial decay width of Ref.\cite{KCH}
--- which is the thrust of our present work.

\bigskip\noindent
For the two-photon decay mode, then, our final results are
\be
\Gamma(\Phi \to \gamma\gamma) = \frac{1}{64\pi} \frac{M_\Phi^3}{\Lambda_\Phi^2} 
\left(\frac{ \alpha}{\pi}\right)^2 \left|I_\gamma\right|^2 
\qquad
{\rm where}
 \qquad I_\gamma = b_{QED} + I_W + \sum_{f} N_c~Q_f^2~I_f
\ee
In the above, $N_c$ is the number of colours of the fermion $f$ and $Q_f$
is the fermionic charge. The QED beta function (appearing because of the
trace anomaly)  is given by\cite{UMA1,UMA2}
\bearr
b_{QED} & = &  \frac{20}{9}  ~{\rm for}~~M_\Phi \leq 2M_W  \nonumber \\
        & = &  \frac{31}{18} ~{\rm for}~~2M_W < M_\Phi \leq 2m_t\nonumber \\
	& = &  \frac{12}{6}  ~{\rm for}~~M_\Phi > 2m_t
\eearr
and the loop integral functions $I_W$ and $I_f$ are given by
\bearr
I_W & = & -1 - \frac{3}{2}\lambda_W + \frac{3}{2}\lambda_W 
(1 - \frac{1}{2}\lambda_W)~{\rm F}(\lambda_W) \nonumber \\
I_f & = & - 8 \lambda_f - \lambda_f ( 4 \lambda_f - 1)~{\rm F}(\lambda_f)
\label{function}
\eearr 
where $\lambda_i = \left({2m_i}/{M_\Phi}\right)^2$, with $i$ running over 
all the particles involved, and ${\rm F}(\lambda)$ is given by
\bearr
{\rm F}(\lambda) 
& = & -2\left[\sin^{-1} \frac{1}{\sqrt{\lambda}}\right]^2 ~~{\rm for}~~ 
\lambda \geq 1 \nonumber \\
& = & -\frac{\pi^2}{2} + \frac{1}{2}\log^2 
\frac{1 + \sqrt{1 - \lambda}} {1 - \sqrt{1 - \lambda}}
- i\pi\log \frac{1 + \sqrt{1 - \lambda}} {1 - \sqrt{1 - \lambda}}
~~{\rm for}~~ \lambda < 1 \nonumber
\eearr
This partial decay width differs from the analogous $H^0 \to
\gamma\gamma$ decay width\cite{BP} in two major particulars, viz.,
\bit
\item The presence of the trace anomaly, i.e. the $b_{QED}$ term.
\item  A different function $I_f$ in Eqn.~(\ref{function}) from that given in, for
example, Ref.\cite{BP}.
\eit

\bigskip\noindent
We now go on to calculate the very-similar process $\Phi \goes gg$, which
yields a partial decay width
\be
\Gamma(\Phi\to gg) = \frac{1}{64 \pi} \frac{M_\Phi^3}{\Lambda_\Phi^2} 
\left(\frac{ \alpha_s}{\pi}\right)^2 \left|I_g\right|^2
\qquad
{\rm where}
\qquad  I_g = b_{QCD} + \sum_{f} \sqrt{2}~I_f
\ee
and, obviously, there is no $I_W$ contribution. The QCD beta function is given
by the usual formula
\bearr
b_{QCD} = 11 - \frac{2}{3}N_f 
& = & \frac{23}{3} ~{\rm for}~~M_\Phi \leq 2M_t \nonumber \\
& = & \frac{21}{3} ~{\rm for}~~M_\Phi > 2M_t 
\eearr
and for $\alpha_s$ we take the usual running value governed by $b_{QCD}$.

\begin{center}
\includegraphics[height=3in]{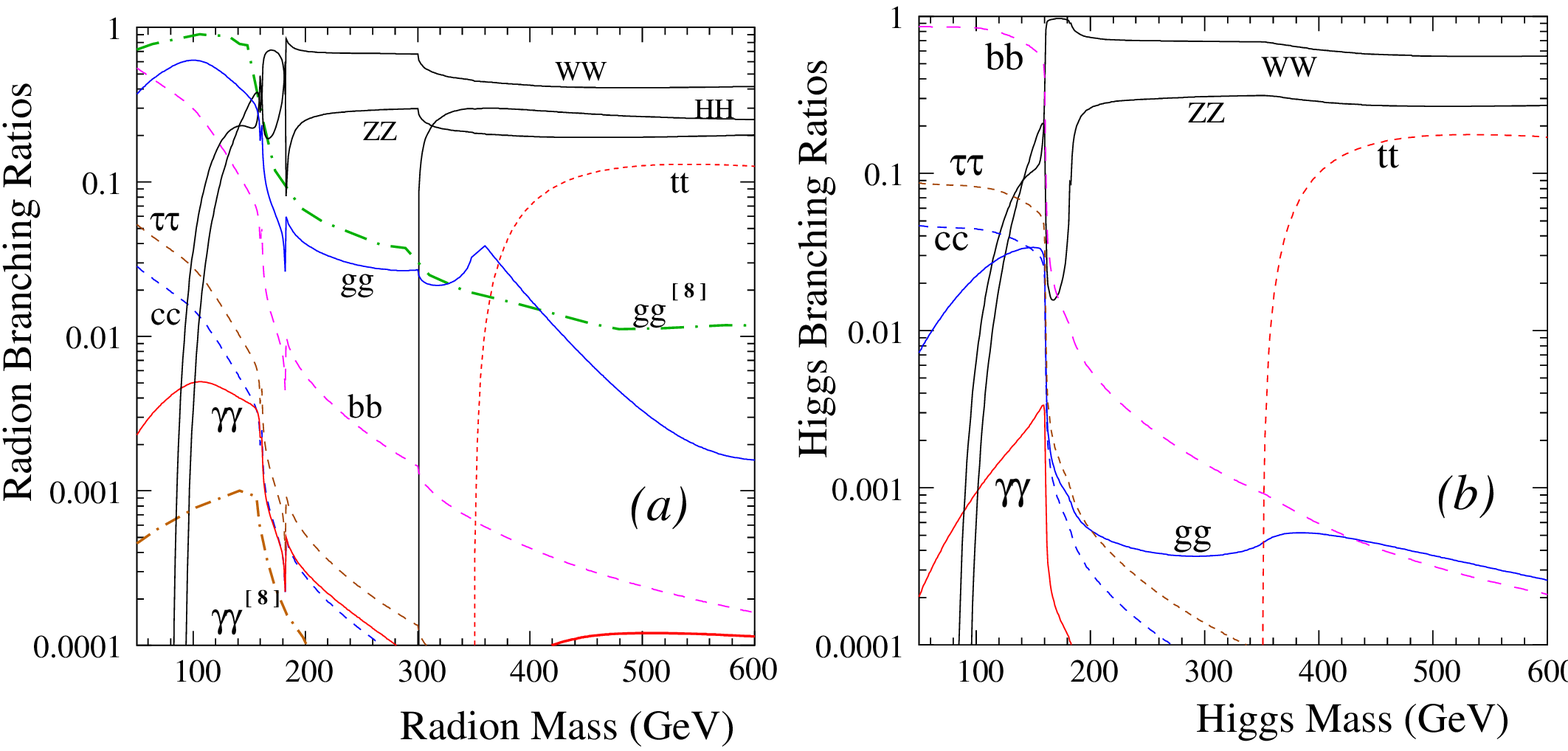}
\end{center}
\vskip -15pt
\noindent {\bf Figure 2}.
{\footnotesize\it Branching ratios for ($a$) a radion and ($b$) a Higgs boson
(of the Standard Model) as a function of the mass. Kinks at kinematic 
thresholds 
are mostly due to numerical instabilities. Note the enormous difference in the 
two-gluon decay mode for the two cases.}

\bigskip\noindent
Branching ratios of the radion to different decay channels may now be
calculated by combining the above formulae with those given in, for
example, Ref.\cite{KCH}, and varying the radion mass $M_\Phi$. Obviously
there will be no dependence on the radion vev, since all the partial
decay widths contain the same factor $\Lambda_\Phi^{-2}$. We have
exhibited our results in Figure~2($a$), which show the principal
branching ratios of the radion, assuming a Higgs boson mass of 150~GeV.
For comparison, Figure~2($b$) shows a similar set of branching ratios
(except the HH decay mode) for a Standard Model Higgs boson with masses
run over the same range. The dot-dashed lines marked $gg^{\cite{KCH}}$
and $\gamma\gamma^{\cite{KCH}}$ correspond to earlier results presented
in Ref.~\cite{KCH} where the momentum dependence of the
radion-fermion-antifermion coupling had not been taken into account. It
may be noted that the difference is quite significant, and indicates that
a cancellation takes place between the finite part of the loop diagram
and the trace anomaly, which is more pronounced for a heavy radion.

\bigskip\noindent
As the figure shows, the decay patterns of the two scalar particles in
question exhibit a great deal of similarity but have some significant
differences also.  Once above the $WW$ and $ZZ$ thresholds, both decay
primarily to weak bosons, with a small percentage of top-anti top decays
when the corresponding threshold is crossed. Both show significant
branching ratios for the $WW^*$ and $ZZ^*$ modes in the scalar mass range
between $M_W$ to $2M_W$ and $M_Z$ to $2M_Z$. At small masses, again, both
show large branching ratios for the $b\bar b$ decay mode, as well as some
for the $c\bar c$ and $\tau^+\tau^-$ channels. However, there the
similarity ends. The radion $\Phi$ has a $\Phi \to HH$ decay channel,
which is obviously forbidden for the Higgs boson. 
Of greater interest is the loop-mediated decay $\Phi \to
\gamma\gamma$, which is at the level of a few per mil when the radion is
light, but is much smaller for a Higgs boson of corresponding mass. (Of
course, such light Higgs bosons have not been found at LEP, so this decay
mode is not really of much use in distinguishing radions from Higgs
bosons.) However, the real {\sl pi\`ece de resistance} is the $\Phi \to
gg$ channel, which has a branching ratio around two orders of magnitude
larger than that of the usual $H \to gg$ process. This branching also
dominates when only the trace anomaly contribution is taken and people
have presented ways of distinguishing radions and Higgs in this
light\cite{huitu}.  Apart from enhancing the branching ratio for a radion
decaying to two hadronic jets significantly above the similar decay of
the Higgs boson, it reduces, (for a light radion, the branching ratio to
a $b\bar b$ pair quite significantly.)  In fact, we find that for light
radions, the {\sl dominant} decay mode is to gluon pairs, while for a
light Higgs boson, the dominant decay mode is to $b \bar b$. Since the
last decay can be pinned down with a fair degree of efficiency by
$b$-tagging methods, we obtain another means of distinguishing between
radions and Higgs bosons.

\bigskip\noindent
As an example of the efficacy of these ideas, we now consider a 1~TeV linear
$e^+e^-$ collider and calculate the production of a radion in association
with a $Z^0$ boson through a process of the form
\bearr
e^+ e^- & \goes           & Z^0 + \Phi \nonumber \\
        & \hookrightarrow & \ell^+ \ell^- + \Phi \nonumber
\eearr
which is then compared with the usual Higgs-strahlung
process\cite{HSTAHL} obtained by replacing $\Phi$ by $H^0$ in the above
process and considering the case when the masses and vev's are equal (or
comparable). In the above $\ell = e, \mu, \tau$, and we have folded in
the relevant detection efficiencies (i.e. 90\% for $\ell = e, \mu$ and
80\% for $\ell = \tau$). Of course, there is a fundamental difference in
the two cases because the electroweak vev $v_{ew} \simeq 246$~GeV is
known, while the radion vev $\Lambda_\Phi$ is an unknown parameter. This
feature is also taken care of in our analysis.

\bigskip\noindent
The discussion in the preceding paragraph makes it clear that it is
interesting to focus on two kinds of final states, viz.
\benum
\item $e^+ e^- \goes \ell^+ \ell^- +$ two jets, which arises when the scalar 
particle decays to a pair of light quarks or gluons\footnote{We exclude 
$\tau^\pm$ decays because these produce narrow jets which can be identified 
as $\tau^\pm$ with 80-90\% efficiency.}, which then undergo fragmentation to 
form a pair of hadronic jets.  
Clearly, for a Higgs boson, this final state will receive contributions
mainly from the decays $H^0 \to b\bar b$ and $H^0 \to c \bar c$, with a
minuscule contribution due to $H^0 \to gg$. However, the radion decay
will have a much larger contribution from the $gg$ mode, and hence the
overall branching ratio to jets should be somewhat higher.

\item $e^+ e^- \goes \ell^+ \ell^- + b \bar b$, which simply means that
the final state in the above contains two tagged $b$-jets. The decay
width for $\Phi \to b \bar b$ is roughly the same as that for $H^0 \to b
\bar b$ when the masses and couplings are the same. However, the presence
of the two-gluon decay mode makes the branching ratio for the $b \bar b$
mode fall quite a bit as compared to that for the Higgs boson when the
radion is light. Of course, the $b \bar b$ cross-section will have to be
convoluted with efficiency factors, which we take\cite{BTAG} to be
$\eta_b = 0.45$ for each tagged $b$-quark.
\eenum
For both kinds of final states, the cross-section will be proportional
to, respectively, $v_{ew}^{-2}$ and $\Lambda_\Phi^{-2}$, and a direct
comparison between the Higgs boson and the radion is meaningful only if
these match, i.e. $\Lambda_\Phi = v_{ew} \simeq 246$~GeV --- a
possibility which, though not ruled out, may not, in general, be
realised. However, if we consider the {\sl ratio} of the two processes,
viz.
$\frac{\sigma(e^+ e^- \goes \ell^+ \ell^- + {\rm two~jets})}
      {\sigma(e^+ e^- \goes \ell^+ \ell^- + b \bar b)}$
the dependence on the vev cancels out and the differences between the two
cases are, therefore, more robust. In fact, the underlying scalar
production process being the same, this ratio is more-or-less equal to
the ratio of the branching fractions $\frac{B(H/\Phi \to {\rm
two~jets)}}{B(H/\Phi \to b \bar b)}$, the only difference being due to
efficiency factors.

\begin{center}
\includegraphics[height=2in]{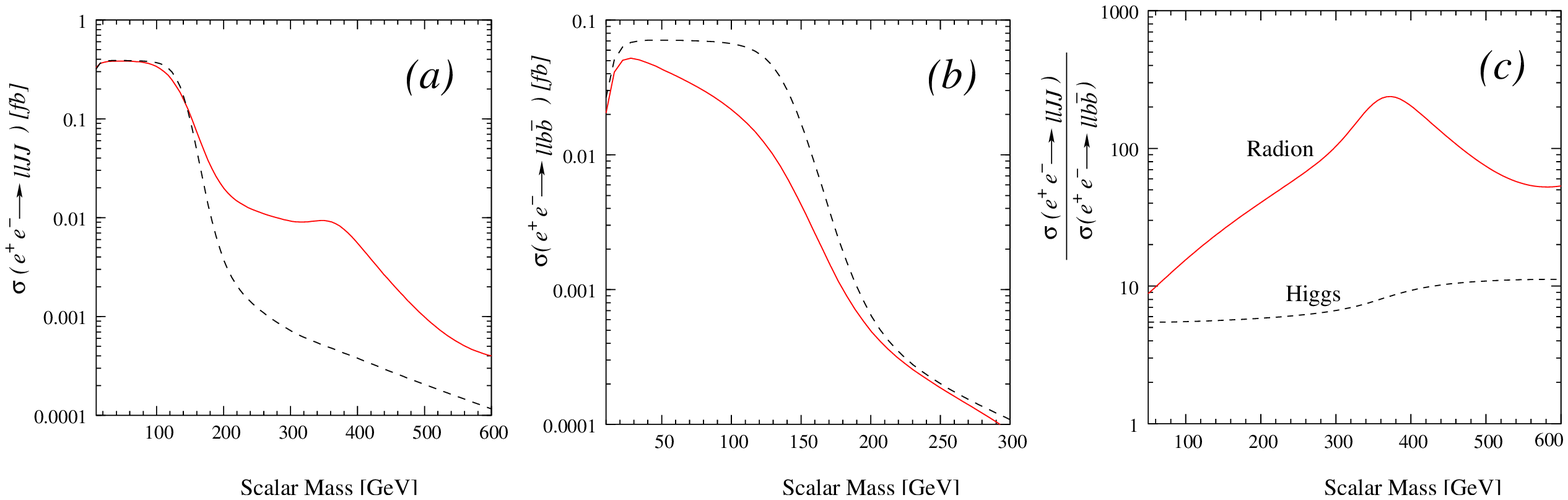}
\end{center}
\noindent {\bf Figure 3}.
{\footnotesize\it Cross sections (in fb) 
for radiative scalar production in association
with $Z^0$ bosons, with scalar decay into ($a$) two jets and ($b$) two tagged 
$b$-jets, as a function of the scalar mass. The solid (red) line denotes the
prediction from a radion and the dashed (black) line that from a Higgs boson.
We set $\Lambda_\Phi = v_{ew} \simeq 246$~GeV, so that radion and Higgs 
production cross-sections match. The ratio of the two cross-sections is shown in
($c$).}

\bigskip\noindent
In Figure~3 ($a$) and ($b$), we illustrate our results for the two processes 
discussed above, namely, \\
\hspace*{1.5in} ($a$) $e^+e^- \goes \ell^+\ell^-$ + two jets, and \\
\hspace*{1.5in} ($b$) $e^+e^- \goes \ell^+\ell^-$ + two $b$-jets. \\
The solid (red) line denotes the radion-mediated cross-section, while the
dashed (black) line denotes the Higgs-mediated one.  In generating the
above curves, we have imposed a few kinematic acceptance cuts on the
final state particles, viz.,
\benum
\item The final-state leptons should have transverse momentum 
$p_T^{(\ell)} > 20$~GeV.
\item The final-state leptons should have pseudo-rapidity 
$\eta^{(\ell)} < 2.0$.
\item The final-state jets should be clearly distinguishable by having their
thrust axes separated by $\Delta R_{JJ} = \sqrt{\Delta \eta_{JJ}^2 + 
\Delta \phi_{JJ}^2} > 0.4$, which is the usual criterion adopted at, 
for example, the LEP and Tevatron colliders.\footnote{In a parton-level 
calculation, this is simply implemented by calculating $\Delta R$ for the 
parent partons, without using a fragmentation algorithm.}
\item The final-state jets should have transverse momentum 
$p_T^{(J)} > 10$~GeV.
\item The final-state jets should have pseudo-rapidity $\eta^{(J)} < 2.0$.
\eenum
The $b$-tagging efficiency has been taken to be 45\%, which is consistent
with the LEP value and is probably a conservative estimate than
otherwise.  It should be noted that the graphs show the excess
cross-section after removing the non-Higgs part of the Standard Model
contributions (such as $e^+e^- \goes ZZ^*$, etc.). These lead to a large
SM four-fermion background, which is, however, easily reducible by
selecting only events corresponding to peaks in the $\ell^+\ell^-$ and
dijet ($b \bar b$) invariant masses. We have not exhibited the background
analysis in this work because we wish to focus on the {\it distinction}
between the two types of scalar resonances, rather than the mere
detection of a scalar resonance (for which several discussions are
already available in the literature).

\bigskip\noindent
A glance at the cross-sections exhibited in Figure 3 ($a$) and ($b$) will
make it clear that for scalar masses well below the $ZZ$-threshold, the
cross-sections for $\ell^+\ell^-$ plus two jets final state are almost
identical in the two cases, but there is considerable difference if we
tag $b$-jets for scalar masses up to at least 150 GeV. Above the
$ZZ$-threshold there is again significant difference in the total
cross-section, which may be attributed to the enhanced decays of the
radion to gluon jets. Noting that the plots are semi-logarithmic in
nature, the deviations between the two cases are quite large.
Interestingly, the two processes complement each other in the sense that
each shows a deviation in the mass region where the other process does
not. This shows up very clearly in Figure 3~($c$), where the ratio of the
two cross-sections is plotted and there is a very large deviation between
the two cases all through the mass range shown. We thus have a simple and
robust method of distinction between production of the two kinds of
scalar particle: simply measure the cross sections for $e^+e^- \goes
\ell^+\ell^-$ + two jets and for $e^+e^- \goes \ell^+\ell^-$ + two
$b$-jets and compute the ratio. A large ratio ($>10$) indicates a radion,
while a smaller ratio ($5 - 10$) indicates a Higgs particle. Of course,
if the radion vev $\Lambda_\Phi$ is very large, the radion effectively
decouples from the Standard Model fields and, though Figure 3 ($c$)
remains unchanged, the radion production cross-sections shown in
Figure~3($a,b$) dwindle accordingly, so that at some stage they become
impossible to measure. This case -- though not improbable -- is not the
point of interest to us in this work.

\bigskip\noindent
To summarise, then, this work consists of two parts. In the first part,
we have correctly computed the decay width of a radion into two photons
or into two gluons using the non-Higgs-like coupling of the radion to
off-shell fermions (specifically to off-shell top quarks, in this case).
This leads to modest changes in the two-photon branching ratios. However,
we predict large changes in the $b\bar b$ branching ratio for light
radions and to the overall dijet branching ratio for heavy radions, both
occurring because of a greatly-enhanced two-gluon decay mode. Using these
results, we compute a `radion-strahlung' process at a 1~TeV linear
$e^+e^-$ collider and show that the processes $e^+e^- \goes \ell^+\ell^-$
+ two jets, with and without $b$-tagging, may be used to distinguish
signals from the on-shell production of a Higgs boson from those arising
from a genuine radion production event. The ratio of the two cases is a
robust method to distinguish between the two cases, even if the radion
vev $\Lambda_\Phi$ is not well-determined.

\bigskip\noindent
Before ending, it needs to be mentioned that we have not taken into
account the possibility\cite{MIX} of {\sl mixing} between the radion and
the Higgs boson.  This is permissible, and, even if precluded at the
tree-level, will be generated by quantum corrections, because the radion
and the Higgs boson carry the same set of gauge quantum numbers. In this
case, however, it is hardly meaningful to talk of the Higgs boson and the
radion separately --- there will just be two scalar states of disparate
masses, and with couplings scaled suitably by a mixing angle $\xi$. Our
work is, therefore, relevant principally in the limit $\xi \to 0$.
However, it is worth mentioning that even if $\xi$ is finite, the
calculation of the decay widths of the new scalar states to a photon pair
or a gluon pair, will be along the lines indicated in the present work,
and hence, our efforts will not have been entirely in vain.

\bq\noindent\small
The authors acknowledge many useful discussions with the late Uma Mahanta, and
wish to express a deep sense of loss at his passing. This work is, therefore,
dedicated to his memory.
\eq
\normalsize


\vfill
\end{document}